\documentclass{article}
\usepackage{spconf}
\usepackage[pdftex]{graphicx}
\usepackage{epstopdf}
\usepackage[utf8]{inputenc} 
\usepackage[T1]{fontenc}    
\usepackage{hyperref}       
\usepackage{url}            
\usepackage{booktabs}       
\usepackage{amsfonts}       
\usepackage{nicefrac}       
\usepackage{microtype}      

\usepackage{graphics, theorem, times, cite}

\usepackage{tikz}

\usepackage{bbm}
\usepackage{mathrsfs}
\usepackage{adjustbox}
\usepackage{makecell}
\usepackage{wrapfig}

\usepackage{amssymb,amsmath,amsfonts,bm,epsfig,theorem}
\usepackage{rotating,setspace,latexsym,epsf,color,dsfont,caption,mathtools}
\usepackage{subfigure}

\usepackage{algorithm}
\usepackage{algorithmic}
\usepackage[shortlabels]{enumitem}

\input{mysymbol.sty}
\usepackage{needspace}

\newcommand{\QED}{\hfill\ensuremath{\blacksquare}}

\def\E{\E}

\newtheorem{theorem}{\hspace{0pt}\bf Theorem}

\title{Unsupervised Learning for Asynchronous  \\ Resource Allocation in Ad-hoc Wireless Networks}
{\name{Zhiyang Wang$^{\dagger}$ \qquad Mark Eisen$^{\star}$ \qquad Alejandro Ribeiro$^{\dagger}$}
\address{${^\dagger}$ Department of Electrical and Systems Engineering, University of Pennsylvania, PA \\
${^\star}$Intel Labs, Hillsboro, OR 
\thanks{Supported by ARL DCIST CRA W911NF-17-2-0181 and Intel Science and Technology Center for Wireless Autonomous Systems. } }
}

\begin{document}
%
\maketitle
\begin{abstract}
We consider optimal resource allocation problems under asynchronous wireless network setting. Without explicit model knowledge, we design an unsupervised learning method based on Aggregation Graph Neural Networks (Agg-GNNs). Depending on the localized aggregated information structure on each network node, the method can be learned globally and asynchronously while implemented locally. We capture the asynchrony by modeling the activation pattern as a characteristic of each node and train a policy-based resource allocation method. We also propose a permutation invariance property which indicates the transferability of the trained Agg-GNN. We finally verify our strategy by numerical simulations compared with baseline methods. 
\end{abstract}
\section{Introduction}
\label{sec:intro}
Wireless communication systems are increasingly facing the challenge of allocating finite and interfering resource over large scale networks with limited coordination between devices. Resource allocation methods, generally speaking, are usually addressed through optimization methods. Because of their non-convex nature in wireless systems, standard centralized resource allocation policies are obtained through heuristic optimization methods \cite{zhang2006stochastic, sangaiah2020iot,gawali2018task}   or data-driven machine learning methods \cite{xu2019energy,sun2017learning, eisen2019learning,lee2019graph, eisen2020optimal,shen2020graph}. The latter case is seeing growing interest due to its applicability in a wide range of application scenarios and lack of reliance on explicit or accurate system modeling.

Resource allocation problems are made more challenging, however, in the decentralized setting where devices are locally selecting resource levels, such as the case in ad-hoc networks. This framework significantly reduces communication overhead and be more robust to single-node failures in large networks. In fully localized methods, devices make their own decision solely with local state measurements. This formulation has been studied both with traditional heuristic methods \cite{chavez1997challenger} as well as machine learning methods \cite{challita2017proactive}. More sophisticated approaches permit limited information exchanges among neighboring network nodes, which can greatly improve performance. Such is the case in the decentralized implementation of the  WMMSE heuristic \cite{shi2011iteratively}, as well as a variety of learning based methods that employ this broader local information structure \cite{lee2019deep, ye2019deep, nasir2019multi, naderializadeh2020resource}.

While most existing decentralized methods consider synchronous algorithms, fully decentralized systems feature different working clocks across nodes in the network. Asynchronous decentralized resource allocation methods have been considered in the context of online optimization approaches  \cite{rajawat2011cross, bedi2018asynchronous}. Alternatively, the learning-based methodologies may be used in which the parameters of a resource allocation policy are pre-trained under asynchronous working conditions offline. 

In this paper we address the asynchronous decentralized wireless resource allocation problem with a novel unsupervised learning policy-based approach. By considering the interference patterns between transmitting devices as a graph \cite{lee2019graph, eisen2020optimal,shen2020graph}, we capture the asynchrony patterns via the activation of the graph edges on a highly granular time scale. From this graph representation of interference and asynchrony, we implement a decentralized learning architecture as the Aggregation Graph Neural Networks (Agg-GNNs) \cite{gama2019convolutional}. The Agg-GNN policy leverages successive local state exchanges to build local embeddings of the global network state at each node, which is then processed through a convolutional neural network to determine resource allocation decisions. Such a policy is trained offline in an unsupervised manner that captures network performance under asynchronous conditions without explicit modeling. We further demonstrate a permutation invariance property of Agg-GNNs that facilitates transference across varying wireless network topologies. Numerical experiments are carried out to verify that our proposed policy outperforms other baseline heuristic methods.

\section{Asynchronous Resource Allocation}
\label{sec:prob}

We consider an ad-hoc wireless system with $m$ transmitters and $n$ receivers with $m\geq n$, such that a {receiver may be paired with multiple transmitters}. We denote the paired receiver of the $i$-th transmitter as $r(i)$. {Due to fast fading phenomenon, the quality of each link changes rapidly over time.} At each time instance $t=1,2,..$, the channel states of all the links can be characterized by a matrix $\bbH(t)\in\reals_{+}^{m\times m}$. Each element $|h_{ij}(t)|:=[\bbH(t)]_{ij}$ represents the link state between transmitter $j$ and receiver $r(i)$ at time slot $t$. We further consider a set of time-varying node states as $\bbx(t) \in \reals^m$, with each entry $x_i(t):=[\bbx(t)]_i$ representing the state of the $i$-th node at time $t$. Both the link states $\bbH(t)$ and node states $\bbx(t)$ are considered as random, drawn from the joint distribution $m(\bbH, \bbx)$.

\begin{figure}[t]
        \centering
        \usetikzlibrary{decorations.pathreplacing,angles,quotes,calc}

\begin{tikzpicture}
\draw (0 ,0) node[anchor=north] {t=0} -- (0,0.1);
\foreach \x in {1,2,...,6} { 
  
  \draw (\x ,0) node[anchor=north] {\x0} -- (\x,0.1);
};
\foreach \x in {0.2,0.4,..., 6.8} { 
  
  \draw (\x ,0) node[anchor=north] {} -- (\x,0.1);
};
\draw (7 ,0) node[anchor=north, yshift=-.1cm] {...} -- (7, 0.1);

\draw[->] (-0.5,0) -- (7.5,0);

\foreach \y in {0.04, 0.09, 0.14}{
\foreach \x in {0,0.1 ,0.4, 0.8, 1.3, 2 , 2.4, 3.4, 4,4.1,5.2,5.3, 6.1, 6.4}{
\draw[color = blue!40, ultra thick] (\x,\y) -- (\x+0.1 ,\y);}
}

\foreach \y in {0.2, 0.25, 0.3}{
\foreach \x in{ 0,0.3 ,0.5, 0.6, 1.4, 1.5,2.1, 2.2 , 2.8, 3.9, 4, 4.8, 5.5,5.6, 6.2, 6.4 }{
\draw[color = red!40, ultra thick] (\x,\y) -- (\x+0.1,\y);}}

\foreach \y in {0.36, 0.41, 0.46}{
\foreach \x in { 0.2 ,0.3,  1.1, 1.2, 2.5, 2.7 , 2.9, 3.1, 3.3 ,4.3, 4.8, 5.1 ,5.4, 6.6, 6.7 }{
\draw[color = green!40, ultra thick] (\x ,\y) -- (\x + 0.1,\y);
}}

\end{tikzpicture}
         \caption{ { Channel and node states change asynchronously in a granular time scale. The colored bars show the working time of three nodes. }}
        \label{fig_timeline}
\end{figure}
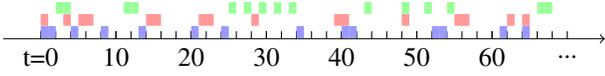

In decentralized networks, it is challenging to maintain a shared synchronous clock across all the nodes without a centralized controller or frequent communication overhead. Figure \ref{fig_timeline} shows an example configuration in which channel and node states vary at each time instance in fast varying scenarios.To design resource allocation policies under asynchronous conditions, we may model heterogeneous and potentially time-varying  ``working patterns'' for each node in the network relative to the more granular channel coherence time given by time index $t=0,1,\hdots$. In particular, at each time $t$, we can denote the set of active nodes as $\ccalA(t)\subseteq \{1,2,\dots, m\}$, which indicates that only nodes $i\in\ccalA(t)$ are capable of making decisions, such as updating its resource allocation strategy, and taking actions, such as sending information to neighboring nodes. 

For decentralized resource allocation policies, it is necessary to define an asynchronous and localized information structure available to each node, which we denote as its \emph{neighborhood}.  In most large-scale network settings, large distance between certain devices causes only negligible channel quality and interference to its receiver. We therefore consider the single-hop neighborhood of node $i$ at time $t$ to consist of active devices with non-negligible interference channel quality, i.e., 
\begin{align}
\label{eqn:neighbor}
\ccalN_{i}(t):=\{j \mid h_{ij}(t) \geq h_{\eps}, j\in\ccalA(t)\}.
\end{align}
In addition to the information received by single-hop neighbors, a device can receive indirect information from larger neighborhoods with some delay to obtain more global state awareness. We define the $k$-hop neighborhood of node $i$ as
\begin{align}
\ccalN_i^k(t) := \{j'\in\ccalN_j(t-k+1), j\in\ccalN_i^{k-1}(t)\cap \ccalA(t) \}.
\end{align}
By imposing a limit of $K$-hop exchanges, the set of local available information of node $i$ at time $t$ is then defined as
\begin{align} 
\label{eqn:local_info}
\ccalH_{i}(t):=\bigcup_{k=0}^{K-1} \Big\{ [\bbH(t-k+1)]_{jj'},& [\bbx(t-k)]_{j'} \mid  \\
&  j\in\ccalN_{i}^{k-1}(t), j'\in\ccalN_{i}^{k }(t) \Big\}. \nonumber
\end{align}
Observe that the local information available at each node depends both on the channel quality with neighboring devices as well as the particular asynchronous activation patterns.

The goal of this work is to find a local resource allocation that maps a node's local history information to an instantaneous resource allocation level $\bbp_i(t) :=\bm\phi_i(\ccalH_i(t); \bbA)$, where $\bbA \in \reals^s$ is the parameter of some function family $\bm\phi(\cdot)$. At time $t$, together with state pairs $\bbH(t), \bbx(t)$, a collection of instantaneous global performance feedbacks $\bbf\left(\bbP(\ccalH(t)), \bbH(t),\bbx(t)\right)$ is experienced in the network. In fast fading channels, users tend to experience the average performance across the states. {This can be evaluated as an expectation over all random states with the assumption that these states are independent and stationary. }The optimal policy then is defined as that which maximizes the total average performance across all the links while respecting local and shared resource budgets $p_0$ and $P_{max}$, respectively, i.e.
\begin{align}
\label{eqn:opt}
\bbA^* := &\argmax_{\bbA, \bbr} \quad  \mathbf{1}^T \bbr,\\
\nonumber & \quad \text{s.t.}\qquad \quad \bbr =\mathbb{E}\left[\bbf\left( \bm{\Phi}(\ccalH,\bbA), \bbH,\bbx\right)\right], \\\nonumber
& \quad \qquad \qquad  \mathbb{E}[\bm{1}^T\bm{\Phi}(\ccalH,\bbA)]\leq P_{max}
,\\\nonumber& \quad \qquad \qquad {\phi}_i(\ccalH_i,\bbA)\in \{0,p_0\}, \quad i=1,..,m.
\end{align}

The optimization problem \eqref{eqn:opt} is typically non-convex and intractable. This limitation and the lack of explicit model-knowledge has motivated the use of model-free unsupervised learning techniques to solve. Numerous decentralized learning architectures have been proposed to solve problems similar to \eqref{eqn:opt}, e.g., fully connected neural networks \cite{lee2019deep, naderializadeh2020resource}. In this work we develop the use of Aggregation Graph Neural Networks (Agg-GNNs) to perform decentralized and scalable resource allocation in asynchronous wireless network settings.

\section{Aggregation Graph Neural Networks}
\label{sec:agg-gnn}
The Agg-GNN resource allocation utilized for decentralized resource allocation can be derived by contextualizing the interference and asynchronous working patterns of the networks as a time-varying graph. Consider a graph with $m$ nodes corresponding to the transmitting devices, and edges $(i,j)$ if $j \in \ccalN_i(t)$. The transmitter states $\bbx(t)$ can be interpreted as signals supported on the nodes, while the instantaneous channel state $h_{ij}(t)$ as the weight of edge $(i,j)$. The weighted graph adjacency matrix is then given by $\bbH_0(t)$, with $[\bbH_0(t)]_{ij}=h_{ij}(t)$ if $j \in \ccalN_i(t)$ and $[\bbH_0(t)]_{ij}=0$ otherwise based on the definition of $\ccalN_i(t)$ in \eqref{eqn:neighbor}. Observe that this graph structure incorporates both the current channel state between devices as well as the asynchronous activation patterns of neighboring devices, and thus represents communicating devices at each time slot.

The Agg-GNN relies on gradually accumulating global state information through an aggregation process in which it collects state information from neighboring transmitters. In particular, node $i$ receives delayed node states $x_j(t-1)$ from its neighbors $j \in \ccalN_i(t)$. This neighborhood information is aggregated by node $i$ into a first hop local feature $y_i^{(i)}(t)$ using its respective channel states $h_{ij}(t)$ as
\begin{equation}\label{eq_transmit_i}
y^{(1)}_i(t) = \sum_{j \in \ccalN_i(t) } h_{ij}(t) x_j(t-1).
\end{equation}
From the definition of $\ccalN_i(t)$, this can be written globally as
\begin{align}
\bby^{(1)}(t) = \bbH_0(t) \bbx(t-1).
\end{align}

The aggregated signal $\bby^{(1)}(t)$ in \eqref{eq_transmit_i} only captures immediate neighborhood state information, but local decisions can be improved relative to the global problem in \eqref{eqn:opt} by accumulating more global information. Thus, aggregated states are further aggregated at the subsequent time $t+1$ under the new adjacency matrix $\bbH_0(t+1)$. After a total of $K$ such successive aggregations, a sequence of aggregated signals $\bby^{(0)}(t), \bby^{(1)}(t), \hdots, \bby^{(K-1)}(t)$ can be obtained, with the $k$-th element for $k >1$ written as
\begin{align} \label{eq:aggregation1}
\bby^{(k)}(t):= \bbH_0(t)\bby^{(k-1)}(t-1) = \left[ \prod_{i=0}^{k-1} \bbH_0(t-i)\right]\bbx(t-k),
\end{align}
while $\bby^{(0)}(t):= \bbx(t)$. By taking the $i$-th element of each element $\bby^{(k)}(t)$, we obtain the \emph{wireless information aggregation sequence} of node $i$, which can be written as:
\begin{align}\label{eq:aggregation2}
\bby_i(t) = [y_i^{(0)}(t) ;y_i^{(1)}(t); \hdots; y_i^{(K-1)}(t)].
\end{align}
This is collected by node $i$ with only the information contained in its local history set defined in \eqref{eqn:local_info}. The aggregated information of each active node is transmitted once per time slot as an overhead message. Note that inactive nodes will not transmit signals but still receive aggregated information, which can then be forwarded to their neighboring nodes during their next active phase. 

\subsection{Graph Neural Networks}


With the local aggregation process in \eqref{eq:aggregation2}, each node obtains a local sequence of state information with a temporal structure. Then it can be processed with a regular Convolution Neural Network (CNN) architecture with $L$ layers.The architecture begins with a standard filter to produce an intermediate output, which is then passed through a pointwise nonlinear function.

For simplicity of presentation, we consider a single feature per layer network structure. We denote $\bm\alpha_l:=[[\bm\alpha_l]_1;\hdots;[\bm\alpha_l]_{K_l}]$ as the coefficients of a $K_l$-tap linear filter which is used to process the feature of the $l-1$-th layer, which is then followed by a pointwise nonlinear function $\sigma_l$. With the initial layer input set as the local aggregation sequence $\bbv_{i0} := \bby_i(t)$,  the $l$-th layer output can be given by
\begin{align}
\bbv_{il}= \sigma_l\left[ \bm\alpha_l * \bbv_{i(l-1)}\right].
\end{align}
The resource allocation is then given by the final layer output, i.e.
\begin{align}
 p_i(t) := \phi_i(\ccalH_{i}(t), \bbA)= \bbv_{iL}.
 \end{align}
The policy parameter is given by the collection of filter coefficients $\bbA=\{\bm\alpha_l \}_{l=1}^L$. Note that the filter parameters are shared across all the nodes, i.e. they implement the same local policy. Although the policy is the same across node, the resource allocations differ for each node due to different local history information. The detailed operation of this process is presented in Alg. \ref{alg:Sequence}.

\begin{algorithm}[h]
\caption{{Resource Allocation at Node $i$}}
\label{alg:Sequence}
\begin{algorithmic}[1]
\FOR{$t=0,1,2..$ }
\IF{Node is active, $i \in \ccalA(t)$}
\STATE{Observe node state $x_i(t)$ and set $y^{(0)}_i(t) = x_i(t)$.}
\STATE{{Transmits sequence $\{y^{(k)}_i(t-1)\}_{k=0}^{K-2}$ to transmitter $j \in \ccalN_i(t)$}.}
\STATE{Node $i$ forms aggregation sequence $\bby_i(t)$ based on information from its active neighbors.}
\STATE{Updates resource level $p_i(t) = \phi(\ccalH_i(t),\bbA) = \bby_{iL}(t)$. }
\ELSE
\STATE{Receives information from its active neighbors and forms aggregation sequence $\bby_i(t)$.}
\STATE{Keeps resource level $p_i(t) = p_i(t-1)$.}
\ENDIF
\ENDFOR
\end{algorithmic}
\end{algorithm}

Observe that the aggregated information sequence in \eqref{eq:aggregation2} is obtained regardless of the dimension and shape of the underlying network---the policy is defined by $\bbA$ and is invariant to input dimension. Furthermore the sequence structure permits invariance to permutations of the corresponding graph. We restrict permutation matrices of dimension $m$ as the permutation set:
\begin{align}
\psi =\{\bm{\Pi} \in\{0, 1\}^{m\times m}: \bm{\Pi}\bm{1}=\bm{1},\bm{\Pi}^T\bm{1}=\bm{1} \}.
\end{align}
The theorem can be stated as follows:
\begin{theorem}
\label{thm:trans}
For any permutation $\bbPi \in \psi$, define the permuted graphs as $\hat\bbH=\bm\Pi^T\bbH\bm\Pi$. The permuted graph signal as: $\hat\bbx=\bm\Pi^T \bbx$. Further define the permuted joint state distribution $\hat{m}(\hbH, \hbx)$ such that $\hat{m}(\hat\bbH,\hat\bbx)=m(\bbH,\bbx)$.
The solutions for \eqref{eqn:opt} $\bbA^*$ and $\hat\bbA^*$ under $\hat{m}(\hat\bbH,\hat\bbx)$ and $m(\bbH,\bbx)$ respectively satisfy $\hat\bbA^*=\bbA^*$.
\end{theorem}

Theorem \ref{thm:trans} indicates we can train an Agg-GNN with state distribution $m(\bbH,\bbx)$ and transfer to a permuted wireless network with state distribution $\hat{m}(\hat\bbH,\hat\bbx)$ without loss of optimality (see \cite{wang2020unsupervised} for proof). This result, along with dimension invariance, suggests potential for transference of an Agg-GNN policy to other networks of varying size, which we study numerically in Section \ref{sec:num}.


\begin{figure*}[t]
\centering
\includegraphics[width=0.32\textwidth]{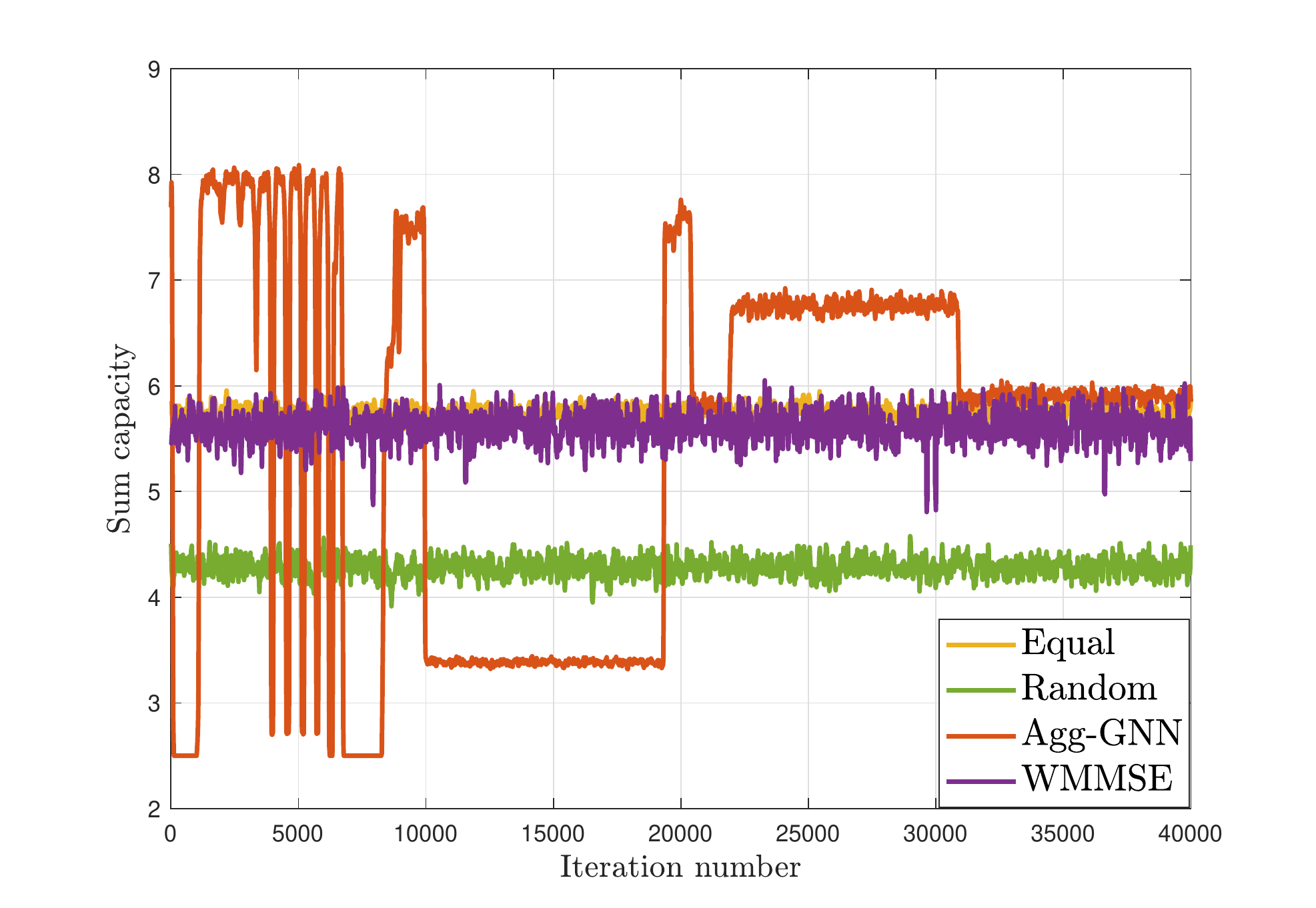}
\includegraphics[width=0.32\textwidth]{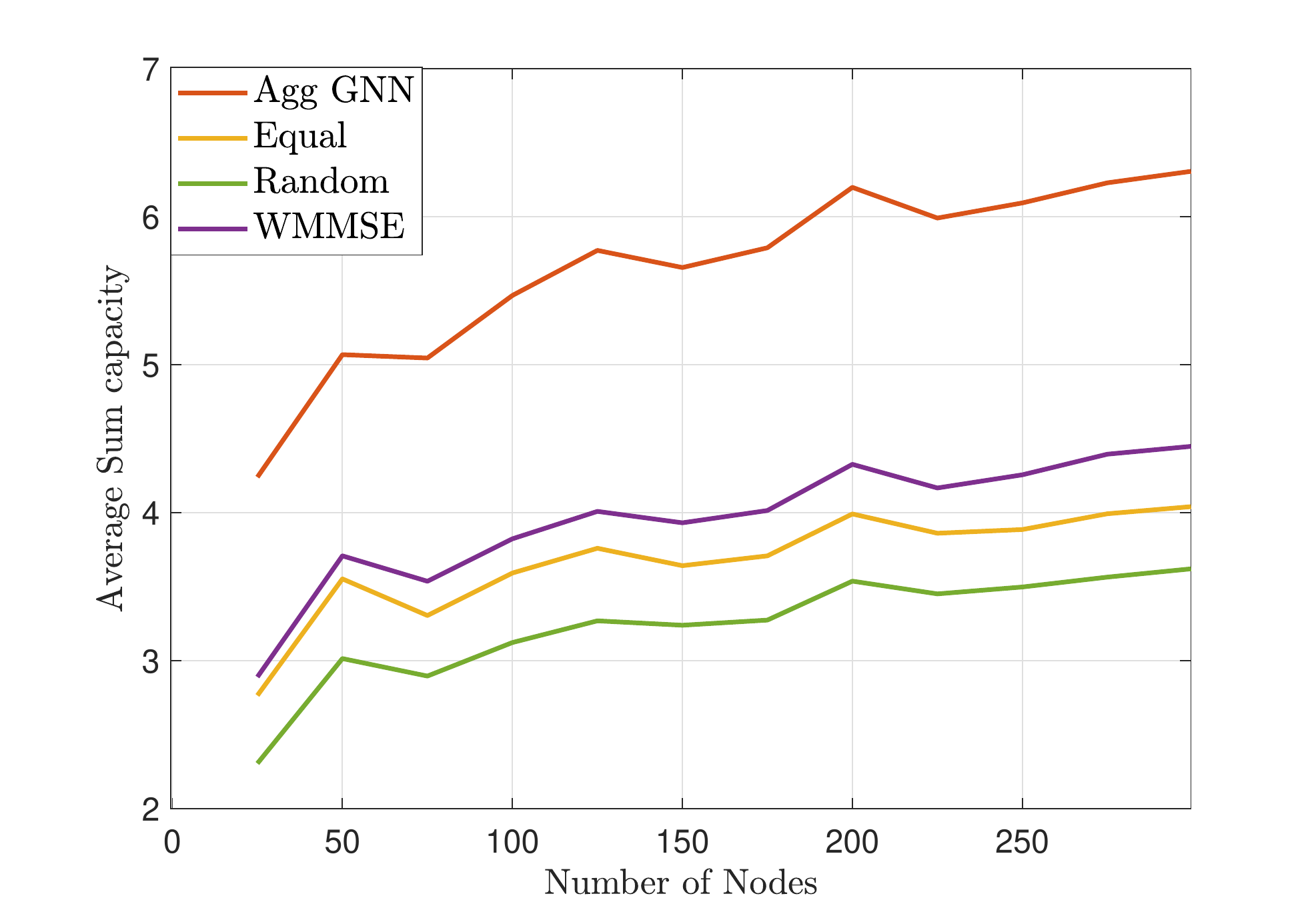}  
\includegraphics[width=0.32\textwidth]{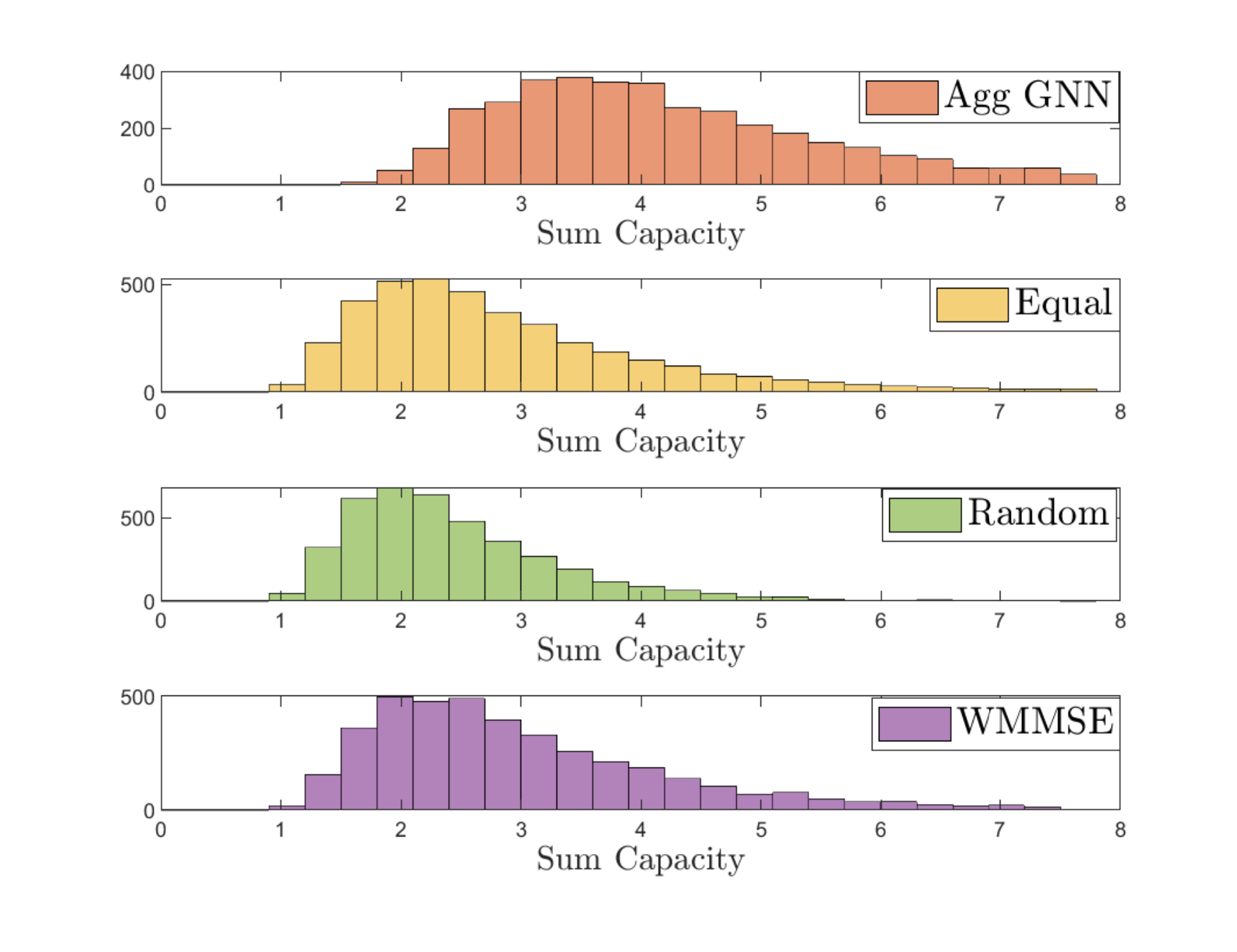}  
\caption{(left) Performance comparison during training for 25 nodes with 5 hops. (middle) Performance comparison when transferring to networks with the same size. (right) Performance comparison when transferring to larger networks. }
\label{fig:trans25}
\end{figure*}
\section{Asynchronous Primal-Dual Training}
\label{sec:p-d}
Finding the optimal GNN filter tensor $\bbA$ requires the solving of a constrained optimization problem in \eqref{eqn:opt}. The constrained optimization problems are often solved by converting to the Lagrangian form, i.e.
\begin{align}
\nonumber\mathcal{L}(\bbA, \bbr, \bm{\lambda}, &\bm{\mu})= \bm{1}^T \bbr + \bm\lambda^T  \left[\mathbb{E}\left[\bbf\left(\bm{\Phi}(\ccalH,\bbA), \bbH,\bbx\right) \right]-\bbr \right] \\
& \qquad + \bm{\mu}^T \left[ \mathbb{E}\left[ \bm{1}^T \bm{\Phi}(\ccalH, \bbA)\right]-P_{max}\bm{1}^T\right]. \label{eq_lagr}
\end{align}
The primal-dual method is used to find the saddle-point of the Lagrangian form by updating primal and dual variables alternatively. Despite having a decentralized network structure, the training of the Agg-GNN parameters is done offline and we can thus leverage a centralized, or federated, learning architecture. This is in contrast to online methods, e.g. \cite{bedi2018asynchronous}, which make resource allocation decisions during a continuous optimization process and thus require decentralized learning.

The primal-dual method is derived by alternating gradient steps on the primal variables $\bbA, \bbr$ and dual variables $\bbmu, \bblambda$. We denote $\tau$ as an iteration index and $\epsilon$ as the stepsize. Although computing gradients of \eqref{eq_lagr} requires explicit knowledge of $\bbf$ and the distribution $m(\bbH,\bbx)$, model-free learning methods such as stochastic gradient and policy gradient methods \cite{sutton2000policy} are typically used to bypass this requirement by sampling the performance under different policies---see \cite{eisen2020optimal}. However, because of the asynchronous working patterns of nodes, not all devices will utilize the current parameter tensor $\bbA(\tau)$ at current index $\tau$. While the centralized learner keeps an iterate $\bbA(\tau)$, we further define the local copy at node $i$, $\bbA_i(\tau)$, and store all together in $\mathbb{A}(\tau) := \{\bbA_i(\tau)\}_{i=1}^m$. The local copies can be expressed as 
\begin{equation}\label{eqn:lcl_A'}
\bbA_i(\tau) := \begin{cases}
\bbA(\tau) & \text{if } i \in \ccalA(\tau), \\
\bbA_i(\tau-1) & \text{if } i \notin \ccalA(\tau).
\end{cases}
\end{equation}

The primal updates with gradient ascent are given by
\begin{align}
\label{eqn:prim1} \bbr(\tau+1)&=\bbr(\tau)+\epsilon [\bm{1} - \bm\lambda(\tau)],\\
\nonumber \bbA(\tau+1) &= \bbA(\tau) + \epsilon  \nabla_{\bbA}\mathbb{E}\left[\bbf\left(\bm\Phi(\ccalH,\mathbb{A}), \bbH,\bbx\right) \right]  \bm\lambda(\tau) \\
\label{eqn:prim2} &\qquad \quad\qquad +\epsilon \nabla_{\bbA}\mathbb{E}\left[ \bm{1}^T \bm\Phi(\ccalH,\mathbb{A}) \right] \bm\mu(\tau)
\end{align}
The dual updates are then likewise given by
\begin{align}
\label{eqn:dual1} \bm\mu(\tau+1) &= \left[\bm\mu(\tau) -\epsilon \left[ \mathbb{E}\left[ \bm{1}^T \bm{\Phi}(\ccalH, \mathbb{A})\right]-P_{max}\bm{1}^T\right] \right]^+,\\
 \bm\lambda(\tau+1)& = \bm\lambda(\tau) - \epsilon\left[\mathbb{E}\left[\bbf\left( \mathbf{\Phi}(\ccalH,\mathbb{A}),  \bbH,\bbx\right) \right]-\bbr(\tau)\right] \label{eqn:dual2} 
\end{align}
We emphasize in \eqref{eqn:prim1}-\eqref{eqn:dual2}, that the model-free evaluations of gradients are taken with respect to the \emph{asynchronized} versions of the policies as given by the paramter tensors $\mathbb{A}(\tau)$, rather than the centralized iterate $\bbA(\tau)$.

%

\section{Simulation Results}
\label{sec:num}
In this section, we provide a numerical study for the problem in \eqref{eqn:opt}. We focus on the wireless ad-hoc networks with $m=n=25$. We first drop $m$ transmitters randomly uniformly within the range of $\bba_i\in[-m,m]^2, i=1,2\dots, m$. Each paired receiver is located randomly within $\bbb_i\in [\bba_i+[-m/4, m/4]]^2$. The fading channel state is composed of a large-scale pathloss gain and a random fast fading gain, which can be written as: $h_{ij}=h^l_{ij}h^f_{ij}$, $h^l_{ij}=\| \bba_i - \bbb_j \|^{-2.2}$, $h^f_{ij}\sim \text{Rayleigh}(2)$. The local and total resource budgets are set as: $p_0=2$ and $P_{max}=m$. We model the active patterns of nodes by considering a collection of $N_{act}=5$ active subsets denoted as $\{\ccalA_n\}_{n=1}^{N_{act}}$, where $\ccalA_n\subseteq  \{1,2,\hdots,m\}$ for all $n$. We set the  number of active nodes at each time step to be a Poisson distributed random variable with $\lambda=12$. At each time $t$, we randomly draw a set of active nodes $\ccalA(t) \in\{\ccalA_n\}_{n=1}^{N_{act}}$. We construct a CNN with $L=10$ hidden layers, each with a filter (hop) with length $K_l=5$ and a standard ReLu non-linear activation function, i.e. $\bm{\sigma}(\bbz)=[\bbz]_+$. The final layer normalizes the outputs through a sigmoid function. 

We compare our algorithm with three existing heuristic methods for solving the original optimization problem: (i) WMMSE \cite{shi2011iteratively} in a distributed setting with limited communication exchanges, (ii) Equal allocation, i.e. assign $P_{max}/m$ to all nodes, and (iii) random allocation, i.e. each node transmits at $p_0$ with probability $P_{max}/(p_0 m)$ to meet the total average power constraint in \eqref{eqn:opt}. Both the proposed Agg-GNN and WMMSE require state information exchange; in comparisons we keep the exchanges complexity equal between these methods, i.e. $K=5$. 


In Fig. \ref{fig:trans25} we evaluate the performance of the Agg-GNN policy and learning method in the asynchronous network. In the left figure, we show the performance relative to baselines during the training procedure on a fixed network and see that, after convergence,  the Agg-GNN narrowly exceeds the performance of the model-based WMMSE method. We next evaluate transference capabilities of the learned Agg-GNN policy on new randomly drawn networks. In the middle figure, we show a histogram of total capacity of the network using the policy in new networks of size 25 compared to baselines. In the right figure, we plot the performance of the learned network and baselines on new networks with growing size. We can see that the proposed Agg-GNN outperforms other heuristic algorithms in all cases. This transference across networks indicates that decentralized strategies for large scale wireless networks can be obtained by training an Agg-GNN on a similar network or a representative smaller network due to their permutation invariance.

\section{Conclusion}
\label{sec:con}
We consider the problem of asynchronous resource allocation in wireless networks.  We train an Aggregation Graph Neural Network with primal-dual model-free unsupervised learning method. Each node aggregates information from its active neighbors with certain delay. The information sequence incorporates the underlying network structure as well as the asynchrony of this system. We propose a policy based on Aggregation Graph Neural Networks with permutation invariance to network structure. The performance of this policy is verified with numerical simulation results compared with heuristic baseline methods.

%


\vfill\pagebreak



\bibliographystyle{IEEEtran}
\bibliography{references-asyn}

\end{document}